\documentclass[11pt, a4paper, onecolumn]{article}

\usepackage[utf8]{inputenc}
\usepackage{graphicx}
\usepackage{caption}
\usepackage{subcaption}
\usepackage{tabularx}
\usepackage{multirow}
\usepackage{array}
\usepackage{hyperref}
\usepackage{amsmath}
\usepackage{amssymb}
\usepackage{soul}
\usepackage[dvipsnames]{xcolor}
\usepackage{float}
\usepackage{mdframed}
\usepackage{authblk}
\usepackage[left=2.5cm,right=2.5cm, top=2cm, bottom=2cm]{geometry}

\begin{document}

\title{Carrier-phase DNS of ignition and combustion of iron particles in a turbulent mixing layer}

\author[1,6]{T. D. Luu}
\author[1]{A. Shamooni}
\author[1]{A. Kronenburg}
\author[2]{D. Braig}
\author[2]{J. Mich}
\author[2]{B.-D. Nguyen}
\author[2]{A. Scholtissek}
\author[2]{C. Hasse}
\author[3]{G. Thäter}
\author[4,5]{M. Carbone}
\author[3]{B. Frohnapfel}
\author[1,6*]{O. T. Stein}

\affil[1]{\small Institute for Combustion Technology, University of Stuttgart, Pfaffenwaldring 31, \newline 70569 Stuttgart, Germany}
\affil[2]{\small Institute for Simulation of Reactive Thermo-Fluid Systems, TU Darmstadt, Otto-Berndt-Stra{\ss}e 2, \newline 64287 Darmstadt, Germany}
\affil[3]{ \small Institute of Fluid Mechanics, Karlsruhe Institute of Technology, Kaiserstra{\ss}e 10, \newline 76131 Karlsruhe, Germany}
\affil[4]{ \small Theoretical Physics I, University of Bayreuth,  Universitätsstra{\ss}e 30, \newline 95447 Bayreuth, Germany}
\affil[5]{ \small Max Planck Institute for Dynamics and Self-Organisation,  Am Fa{\ss}berg 17, \newline 37077 Göttingen, Germany}
\affil[6]{ \small Engler-Bunte-Institute, Simulation of Reacting Thermo-Fluid Systems, Karlsruhe Institute \newline of Technology, Engler-Bunte-Ring 7, 76131 Karlsruhe, Germany}
\affil[*]{\small Corresponding author: oliver.t.stein@kit.edu}

\date{January 2024}
\maketitle

\begin{abstract}
Three-dimensional carrier-phase direct numerical simulations (CP-DNS) of reacting iron particle dust clouds in a turbulent mixing layer are conducted. The simulation approach considers the Eulerian transport equations for the reacting gas phase and resolves all scales of turbulence, whereas the particle boundary layers are modelled employing the Lagrangian point-particle framework for the dispersed phase. The CP-DNS employs an existing sub-model for iron particle combustion that considers the oxidation of iron to FeO and that accounts for both diffusion- and kinetically-limited combustion. At first, the particle sub-model is validated against experimental results for single iron particle combustion considering various particle diameters and ambient oxygen concentrations. Subsequently, the CP-DNS approach is employed to predict iron particle cloud ignition and combustion in a turbulent mixing layer. The upper stream of the mixing layer is initialised with cold particles in air, while the lower stream consists of hot air flowing in the opposite direction. Simulation results show that turbulent mixing induces heating, ignition and combustion of the iron particles. Significant increases in gas temperature and oxygen consumption occur mainly in regions where clusters of iron particles are formed. Over the course of the oxidation, the particles are subjected to different rate-limiting processes. While initially particle oxidation is kinetically-limited it becomes diffusion-limited for higher particle temperatures and peak particle temperatures are observed near the fully-oxidised particle state. Comparing the present non-volatile iron dust flames to general trends in volatile-containing solid fuel flames, non-vanishing particles at late simulation times and a stronger limiting effect of the local oxygen concentration on particle conversion is found for the present iron dust flames in shear-driven turbulence.
\end{abstract}

\section{Introduction}
\label{sec:int}

Due to industrial growth in both developing and developed countries, as well as the increasing world population, the global energy demand is expected to rise over the next few decades. Energy from fossil fuels is still dominant, but their negative environmental impact and limited resources have raised the need for alternative energy sources. Renewable energy sources from wind and solar are clean (carbon-free) but due to  the geographical locations of wind farms and volatility from solar plants, regions with high energy demand cannot be supplied continuously. A possible approach to solve this issue is the use of carbon-free energy storage systems. Nowadays, energy storage technologies are still underdeveloped and partially unexplored. A promising emission-free technology for future energy systems is the oxidation of iron \cite{bergthorson2015, debiagi2022} and the potential of using existing coal power plant infrastructures \cite{janicka2023}. Iron has a high potential for serving as a carbon-free energy carrier due to its high energy density and abundance combined with its excellent transport and storage properties \cite{bergthorson2015, bergthorson2018}. When combining iron oxidation with the reverse iron oxide reduction process, based on renewable energy sources, a sustainable circular zero-carbon energy economy can be developed \cite{julien2017, bergthorson2018}. Due to its non-volatile heterogeneous combustion property \cite{goroshin2022} iron combustion differs significantly from coal and biomass conversion such that existing modelling strategies for carbon-based solid fuels cannot directly be used. This lack of reliable modelling approaches drives significant research efforts to examine the underlying chemical and physical processes, and to develop new models.

Goroshin et al. \cite{goroshin1996} developed a simple analytical model for single metal particles to capture heating, ignition and diffusive burnout in both the fuel-lean and fuel-rich limits. Extension of the analytical model for binary mixtures of reacting fuels has been conducted by Goroshin et al. \cite{goroshin2000} and Palecka et al. \cite{palecka2018}. Independently, Sidorov and Shevchuk \cite{sidorov2011} developed a similar analytical model. Goroshin et al. \cite{goroshin2011}, Tang et al. \cite{tang2012} and Lam et al. \cite{lam2017} developed numerical models to investigate the discrete combustion regime that may occur for solid metal fuels. Soo et al. \cite{soo2015} proposed a kinetically- and diffusion-limited model for iron that was later extended by Hazenberg and van Oijen \cite{hazenberg2021}, considering the major oxidation step of Fe to FeO. Recently, Thijs et al. \cite{thijs2022} and Mich et al. \cite{mich2023} extended this model by deriving improved heat and mass transfer correlations from fully-resolved particle simulations \cite{thijs2022} and studying polydispersity effects on single particle combustion in the Euler-Lagrange framework \cite{mich2023}. Mi et al. \cite{mi2022} proposed an alternative iron combustion model that describes the growth of iron oxide layers consisting of FeO and Fe$_3$O$_4$ by a parabolic rate law. Over the last couple of years, the step from single particle investigations \cite{ning2021, ning2022, ning2022b, li2022, liT2022, fujinawa2023, panahi2023, ning2023} to the study of metal particle clouds in laminar flows has received a substantial boost both numerically and experimentally. The experimental investigations mainly focus on the stabilisation and quantification of laminar burning velocities and on identifying different combustion regimes \cite{tang2009, goroshin2011b, tang2011, julien2015, wright2016, mcrae2019, palecka2019, poletaev2022, fedoryk2023}. Numerical studies are based on single particle model developments and examine various physical aspects such as discreteness \cite{goroshin2011, tang2012, mi2017, lam2017}, flame structure and laminar burning velocities \cite{palecka2018, goroshin2000, hazenberg2021, ravi2023, wen2023}.

The transition from laminar to turbulent iron cloud combustion introduces additional challenges. From previous research on solid fuel combustion with volatiles, it is known that a strong coupling between turbulent mixing, homogeneous chemistry and solid fuel kinetics exists. However, the non-volatile nature of iron requires us to re-examine this interplay for iron cloud flames. Direct numerical simulations (DNS) are essential to investigate these processes in detail. While fully-resolved DNS, that resolves all particle boundary layers, is only possible for single particles \cite{thijs2022} and small particle groups, the carrier-phase direct numerical simulation (CP-DNS) approach provides a good trade-off between accuracy and efficiency for particle cloud combustion. CP-DNS resolves all scales of turbulence and the flame, but employs sub-models for the momentum, heat and mass transfer across the boundary layers between the bulk gas phase and the Lagrangian point particles. Recent CP-DNS of solid fuel particle cloud combustion with volatile gases from coal or biomass were conducted by several researchers \cite{hara2015, brosh2015, bai2016, rieth2018, wan2019, wen2020e, wang2021, shamooni2021, chen2023}. Hemamalini et al. \cite{hemamalini2023} considered reacting iron particle clouds in a double shear layer configuration by CP-DNS and provided a qualitative impression of early-time iron cloud combustion. Here, we conduct CP-DNS of a reacting iron particle cloud in a turbulent mixing layer by using the First Order Surface Kinetics (FOSK) iron combustion sub-model proposed by Mich et al. \cite{mich2023}. Our objectives are to
\begin{itemize}
 \setlength\itemsep{-0.5mm}
 \item provide a first-of-its-kind study of iron particle cloud ignition and combustion in shear-driven turbulence using existing iron combustion sub-models, 
 \item investigate the non-volatile combustion behaviour of iron particle clouds and qualitatively
compare it to volatile-containing solid fuel flames,
 \item provide a CP-DNS database for further model development of iron particle cloud flames.
\end{itemize}

\noindent
The remainder of the paper describes the modelling approach in Sec.\,\ref{sec:mod}, shows validation results for the iron combustion sub-model in Sec.\,\ref{sec:val} and introduces the computational configuration in Sec.\,\ref{sec:com}. The CP-DNS results are presented and discussed in Sec.\,\ref{sec:res}, followed by conclusions in Sec.\,\ref{sec:con}.

\section{Modelling}
\label{sec:mod}

\subsection{Gas phase}

In the context of CP-DNS, the gas phase is described by the governing equations of mass (Eq. \ref{eq:mass}), momentum (Eq. \ref{eq:momentum}), energy (Eq. \ref{eq:energy}) and chemical species (Eq. \ref{eq:species}) \cite{rieth2018}
\begin{equation}
\frac{\partial \rho}{\partial t} + \frac{\partial}{\partial x_i} (\rho u_i) = \dot{S}_{\rho, p},
\label{eq:mass}
\end{equation}
\vspace{-0.8cm}
\begin{align}
\frac{\partial \rho u_i}{\partial t} &+ \frac{\partial}{\partial x_j} (\rho u_i u_j) = - \frac{\partial p}{\partial x_i} \nonumber \\ 
&+ \frac{\partial}{\partial x_j} \left( \mu \left[\frac{\partial u_j}{\partial x_i} + \frac{\partial u_i}{\partial x_j} - \frac{2}{3} \frac{\partial u_m}{\partial x_m} \delta_{ij} \right] \right) + \dot{S}_{u, p},
\label{eq:momentum}
\end{align}
\begin{equation}
\frac{\partial \rho h_s}{\partial t} + \frac{\partial}{\partial x_i} (\rho u_i h_s) = \frac{\partial}{\partial x_i} \left(\frac{\mu}{\mathrm{Pr}} \frac{\partial h_s}{\partial x_i} \right) + \dot{S}_{h_s, p} + \dot{S}_\mathrm{rad},
\label{eq:energy}
\end{equation}
\vspace{0.2cm}
\begin{equation}
\frac{\partial \rho Y_k}{\partial t} + \frac{\partial}{\partial x_i} (\rho u_i Y_k) = \frac{\partial}{\partial x_i} \left(\frac{\mu}{\mathrm{Sc}} \frac{\partial Y_k}{\partial x_i} \right) + \dot{\omega}_k + \dot{S}_{k,p}
\label{eq:species}
\end{equation}
\noindent
with density $\rho$, time $t$, spatial coordinate $x$, coordinate directions $i$,$j$, velocity $u$, Kronecker delta $\delta_{ij}$ and pressure $p$. The symbol $\mu$ denotes dynamic viscosity, $h_s$ sensible enthalpy, $Y_k$ mass fraction of species $k$ and Pr\,=\,Sc\,=\,0.7 are the non-dimensional Prandtl and Schmidt numbers. $\dot{S}_\mathrm{rad}$ denotes the radiative source term. Radiative heat transfer is modelled by the discrete ordinate method (DOM) using 80 directions. However, assuming a mixture of $\mathrm{O_2}$ and $\mathrm{N_2}$ (pure air) the gas is nearly transparent in the infrared wavelength and does not absorb or emit major parts of the radiation ($\varepsilon_\mathrm{gas} \approx 0.001$). The further RHS source terms $\dot{S}_{\rho, p}$, $\dot{S}_{u, p}$, $\dot{S}_{h_s,p}$ and $\dot{S}_{k,p}$ denote the exchange of mass, momentum and energy between the gas phase and the iron particles, and $\dot{\omega}_k$ is the homogeneous chemical reaction rate of species $k$. Assuming purely non-volatile iron conversion, the iron oxide formed during oxidation entirely remains on the particle in solid form and no homogeneous reactions occur such that $\dot{\omega}_k = 0$. However, the mass transfer terms $\dot{S}_{\rho, p}$ and $\dot{S}_{k,p}$ for $k = \mathrm{O_2}$ are non-zero, since oxygen is consumed from the gas phase. The non-zero source terms are calculated as
\begin{equation}
\dot{S}_{\rho, p} = \dot{S}_{\mathrm{O_2},p} = -\frac{1}{\Delta^3} \sum^{N_p}_{p=1} \frac{dm_p}{dt},
\label{eq:source_mass}
\end{equation}
\begin{equation}
\dot{S}_{u, p} = -\frac{1}{\Delta^3} \sum^{N_p}_{p=1} \frac{d (m_p \textbf{u}_p)}{dt},
\label{eq:source_mom}
\end{equation}
\begin{equation}
\dot{S}_{h_s, p} = -\frac{1}{\Delta^3} \sum^{N_p}_{p=1} \left( \frac{m_p c_{\mathrm{p},p}}{\tau_{con}}(T_g-T_p) + \frac{dm_p}{dt} h_{s, \mathrm{O_2}}\vert_{T_p} \right)
\label{eq:source_ent}
\end{equation}
\noindent
with the particle mass $m_p$, particle temperature $T_p$, gas phase temperature $T_g$ at the particle position, convective heat transfer time scale $\tau_{con}$, particle specific heat capacity $c_{\mathrm{p},p}$ and the sensible enthalpy of consumed oxygen during the oxidation process at particle temperature $h_{s, \mathrm{O_2}}\vert_{T_p}$. Equations \eqref{eq:source_mass}-\eqref{eq:source_ent} specify the semi-discretised source terms from $N_p$ iron particles within every gas phase cell with edge length $\Delta$ and Eq.\,\eqref{eq:source_mom} excludes external forces, e.g. gravity.

\subsection{Solid phase}

The particles are described in the Lagrangian framework and their interaction with the carrier-phase is defined by the transfer of momentum, energy and oxygen mass. The particles are initialised as pure iron. During the oxidation process FeO is produced according to 
\begin{equation}
\mathrm{Fe + 0.5 O_2 \rightarrow FeO}.
\label{eq:reaction} 
\end{equation}
\noindent
Although higher oxidation states of iron (Fe$_3$O$_4$ and Fe$_2$O$_3$) exist, we limit ourselves to the sole production of FeO. This is because a recent set of iron combustion sub-models is based on the same assumption \cite{hazenberg2021,thijs2023,mich2023,wen2023} such that limiting ourselves to FeO results in a better comparability of our data to these recent references. These existing FeO sub-models represent the most important oxidation physics (kinetic and/or diffusion limitation) and have been successfully validated against experimental data. Despite the proven existence of higher oxidation states, there are currently no reliable sub-models that can represent the entire physics of their production for the present conditions. Moreover, with our current focus on ignition, particle residence times in the mixing layer remain short, such that substantial iron oxidation to the final state Fe$_2$O$_3$ does not occur. During the simultaneous presence of both Fe and FeO, their thermophysical properties ($\rho$, $c_p$ and $h_s$) need to be defined. In this work, the densities are calculated in analogy to \cite{mich2023}. The remaining thermophysical properties are taken from the NIST database \cite{chase1998} and described by the Shomate equations. The particle specific heat capacity is expressed as $c_{\mathrm{p},p} = Y_\mathrm{Fe} c_{\mathrm{p, Fe}} + Y_\mathrm{FeO} c_{\mathrm{p, FeO}}$. Melting and solidification of both iron and iron oxide are considered by the apparent heat capacity method \cite{thijs2023}. The mass conversion rate of Fe to FeO is predicted using the FOSK model by Mich et al. \cite{mich2023}
\begin{equation}
 \frac{d{m}_{p, \mathrm{Fe}}}{dt} = - \frac{1}{s} \rho_f Y_\mathrm{O_2} A_d k_d \mathrm{Da}^* ,
\end{equation}
\begin{equation}
 \frac{d{m}_{p, \mathrm{FeO}}}{dt} = \frac{1+s}{s} \rho_f Y_\mathrm{O_2} A_d k_d \mathrm{Da}^*.
 \label{eq:mFeO}
\end{equation}
Here, 
\begin{equation}
\mathrm{Da}^* = \frac{A_r k_r}{A_r k_r + A_d k_d}
\label{eq:damkohler}
\end{equation} 
\noindent
is the normalised Damköhler number, $A_r = A_d = \pi d_p^2$ the reactive and diffusive areas of the particle, $k_r = k_\infty \mathrm{e}^{-E_a/R_u T_p}$ the kinetic surface reaction rate and $k_d = \mathrm{Sh} \frac{D_{\mathrm{O_2}, f}}{d_p}$ the diffusive transfer rate. $\mathrm{Sh} = 2 + 0.552 \mathrm{Re}_p^{1/2}\mathrm{Sc}^{1/3}$ denotes the Sherwood number, $\mathrm{Re}_p = \rho_f | \textbf{u}_g - \textbf{u}_p | \frac{d_p}{\mu_f}$ the particle Reynolds number and $s$ the stoichiometric ratio of the oxidation of iron to iron oxide. Subscript $f$ denotes film properties evaluated using the 1/3-law $T_f = T_p + \frac{1}{3}(T_g-T_p)$. The remaining coefficients in the model are based on \cite{mich2023} and further model details can be found therein.

Only drag force is considered to act on the particles. While it has been found that gravity may have an impact on laboratory scale iron particle flames \cite{fedoryk2023}, it is neglected for the present canonical mixing layer configuration. Uniform temperature is assumed within the particles (\mbox{$\mathrm{Bi} \approx 0.001-0.01$} \cite{mi2022}) which is governed by convective heat transfer with the gas phase, radiation, heat of oxidation and the consumption of oxygen from the gas phase. With these assumptions, the solid phase governing equations read
\begin{equation}
\frac{d m_p}{dt} = \frac{d{m}_{p, \mathrm{Fe}}}{dt} + \frac{d{m}_{p, \mathrm{FeO}}}{dt},
\end{equation}
\begin{equation}
\frac{d \textbf{u}_p}{dt} = \frac{\textbf{u}_g - \textbf{u}_p}{\tau_p},
\end{equation}
\begin{equation}
\frac{d T_p}{dt} = \frac{1}{\tau_{con}} (T_g - T_p) + \frac{\varepsilon_p A_p \sigma}{m_p c_{p, p}} (\Theta^4_r - T_p^4) + \frac{\dot{Q}_\mathrm{FeO}}{m_p c_{p, p}} + \frac{\dot{Q}_\mathrm{O_2}}{m_p c_{p, p}}
\label{eq:particle_temperature}
\end{equation}
\noindent
where $\textbf{u}_g$ is the gas velocity at the particle position, $\textbf{u}_p$ the particle velocity and $\tau_p$ the particle relaxation time that reads 
\begin{equation}
\tau_p =  \frac{\rho_p d_p^2}{18 \mu_f} \frac{1}{(1+0.15 \mathrm{Re}_p^{2/3})}.
\end{equation}
\noindent
$\varepsilon_p = 0.9$ \cite{burgess1916} is the particle emissivity, $\sigma$ the Stefan-Boltzmann constant, $\Theta_r$ the gas phase radiation temperature computed by DOM, $\dot{Q}_\mathrm{FeO} = \frac{dm_{p,\mathrm{FeO}}}{dt}\Delta h_{c, \mathrm{FeO}}$ the heat release due to $\mathrm{Fe}$ oxidation, $\Delta h_{c, \mathrm{FeO}}$ the formation enthalpy of $\mathrm{FeO}$ based on Eq. (\ref{eq:reaction}), and $\dot{Q}_\mathrm{O_2}$ the energy transfer due to oxygen consumption at particle temperature. The convective heat transfer time scale follows Ranz-Marshall \cite{ranz1952}
\begin{equation}
\tau_{con} = \frac{1}{6} \frac{\mathrm{Pr}}{\mathrm{Nu}} \frac{c_{\mathrm{p}, p}}{c_{\mathrm{p},f}} \frac{\rho_p d_p^2}{\mu_f} 
\label{eq:tau_con}
\end{equation}
with
\begin{equation}
\mathrm{Nu} = 2 + 0.552 \mathrm{Re}_p^{1/2}\mathrm{Pr}^{1/3}
\label{eq:Nu}
\end{equation}
\noindent
A Sherwood and Nusselt number correction is applied to account for Stefan flow \cite{spalding1979,bird2002,thijs2022}
\begin{equation}
\mathrm{Sh}^* = \mathrm{Sh} \frac{\mathrm{ln}(1+B_M)}{B_M},
\end{equation}
\begin{equation}
\mathrm{Nu}^* = \mathrm{Nu} \frac{\mathrm{ln}(1+B_T)}{B_T}
\end{equation}
\noindent
such that $\mathrm{Sh}^*$ replaces $\mathrm{Sh}$ in the expression for $k_d$ and $\mathrm{Nu}^*$ is used instead of $\mathrm{Nu}$ in Eq. (\ref{eq:tau_con}). The Spalding mass transfer number $B_M$ is calculated as
\begin{equation}
B_M = \frac{Y_{\mathrm{O_2},g}-Y_{\mathrm{O_2},p}}{Y_{\mathrm{O_2},p}-1} \quad \mathrm{with} \quad Y_{\mathrm{O_2},p} = Y_{\mathrm{O_2},g} \frac{A_d k_d}{A_r k_r + A_d k_d}
\end{equation}
and the Spalding heat transfer number $B_T$ as
\begin{equation}
B_T = (1 + B_M)^{\varphi}-1 \quad \mathrm{with} \quad \varphi = \frac{c_{\mathrm{p},\mathrm{O_2}}\vert_{T_p}}{c_{\mathrm{p},g}} \frac{\mathrm{Pr}}{\mathrm{Sc}}
\end{equation}
\noindent
where $c_{\mathrm{p},\mathrm{O_2}}\vert_{T_p}$ is the specific heat capacity of oxygen at particle temperature and $c_{\mathrm{p},g}$ the specific heat capacity of the gas phase.

\section{Validation}
\label{sec:val}

\begin{figure}[h!]
\centering
\includegraphics[width=0.7\textwidth]{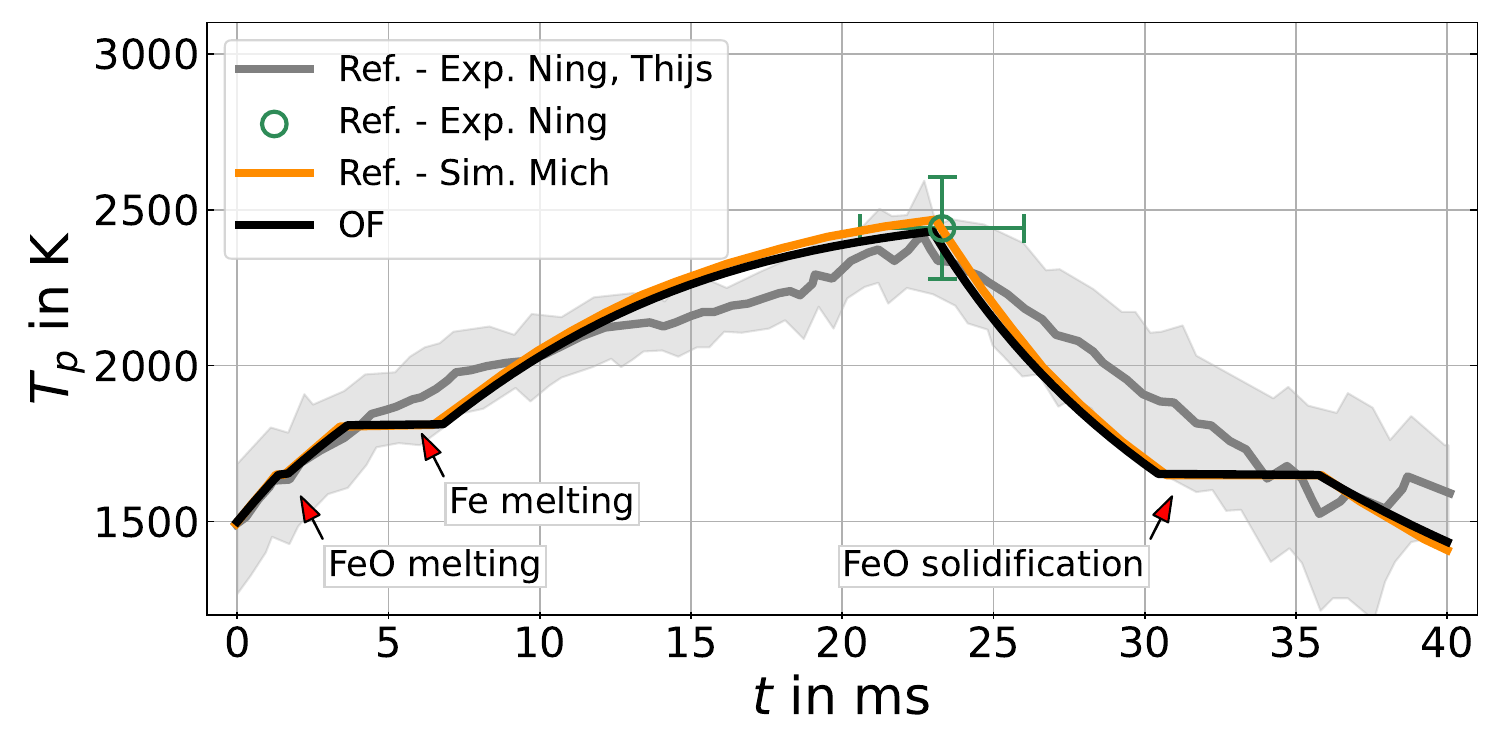}
\caption{Comparison of particle temperature $T_p$ vs. time $t$ for laser-ignited single iron particles with $d_p=54$\,\textmu m in air at $T_\mathrm{air} = 300$\,K. Present work (OF), ref.\,\cite{mich2023} (Ref.--Sim.) and refs\,\cite{ning2021,thijs2023} (Ref.--Exp.) with mean (dark grey line) and standard deviation (gray shading) from the measurements.}
\label{fig:validation_Tp_vs_t}
\end{figure}

The sub-model for iron combustion is validated by comparing results from it for a laser-heated iron particle configuration to numerical predictions from \cite{mich2023} and experimental data published in \cite{ning2021,thijs2023}. The configuration is the laser-heated reactive single iron particle for various oxygen environmental conditions and varying particle diameters by Ning et al. \cite{ning2021}. Figure\,\ref{fig:validation_Tp_vs_t} shows the time evolution of particle temperature for a laser-ignited iron particle with $d_p \approx 54$\,\textmu m in air at $T_\mathrm{air} = 300$\,K.
The initial particle temperature is $T_p = 1500$\,K due to heating by the laser. The laser initiates the heterogeneous reaction, which leads to a steady increase of the particle temperature until $t = 23$\,ms. The particle reaches a peak temperature of $\approx\,2400 $\,K that levels off to $\approx\,1400$\,K at $t = 40$\,ms due to the full consumption of Fe. The time evolution profile exhibits three plateaus of constant particle temperature which are induced by the melting and solidification phenomena inside the iron particle. The first ($1.2$\,$<$\,$t$\,$<$\,$1.6$\,ms) and third plateau ($30.5$\,$<$\,$t$\,$<$\,$36$\,ms) at $T_p = 1650$\,K correspond to the melting and solidification temperature of FeO, while the second plateau ($3.5$\,$<$\,$t$\,$<$\,$6.8$\,ms) at $T_p = 1810$\,K represents the melting temperature of Fe. The first plateau (melting of FeO) is comparatively short, as only a small fraction of FeO has formed during the initial stages of the oxidation process. After the peak particle temperature has been reached at $23$\,ms the particle only consists of FeO and hence, there is no solidification plateau for Fe at $1810$\,K, but only the one for FeO at $1650$\,K during the time period $30.5$\,$<$\,$t$\,$<$\,$36$\,ms. Comparing the results between the two different codes \cite{mich2023} and the experimental data from \cite{ning2021,thijs2023} in Fig.\,\ref{fig:validation_Tp_vs_t}, consistent numerical predictions are observed which both lie within the experimental scatter.

\begin{figure}[h!]
\centering
\includegraphics[width=0.65\textwidth]{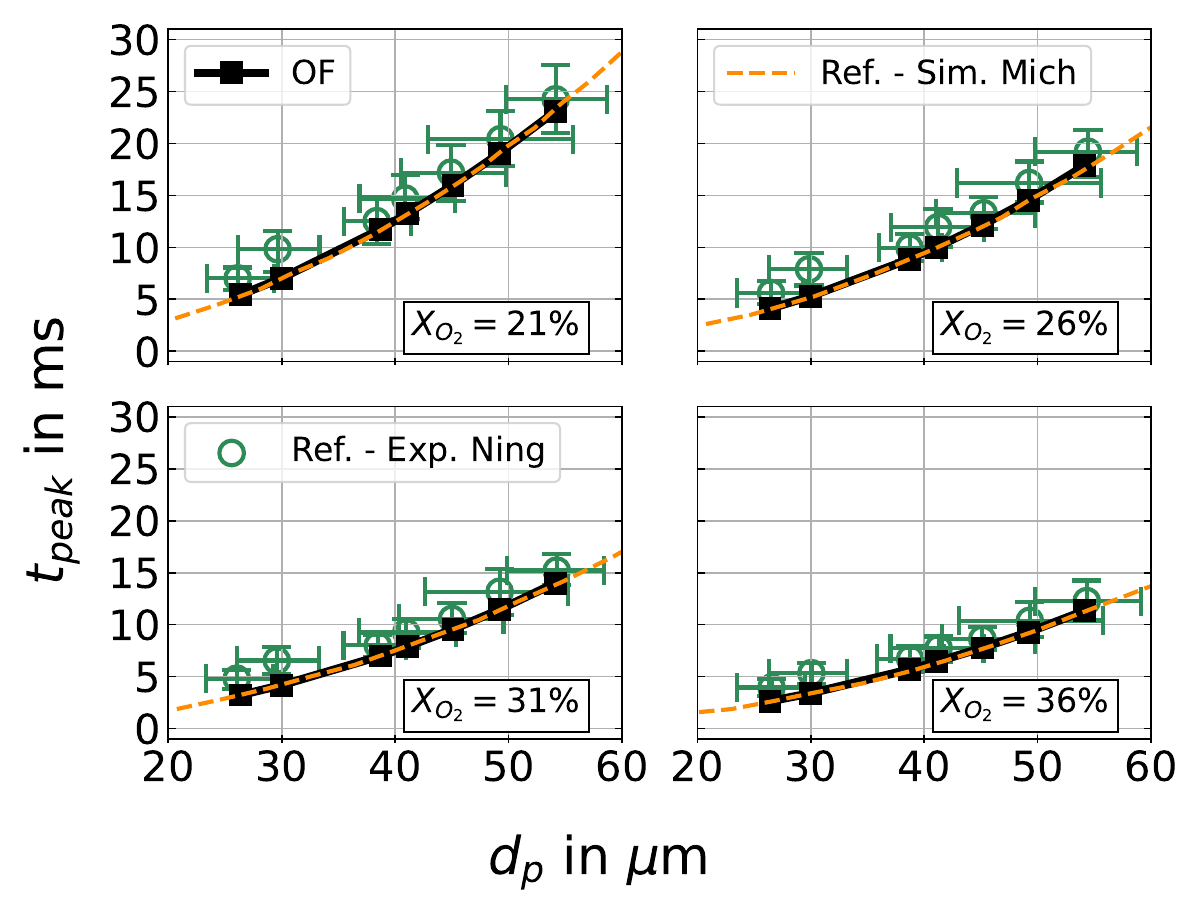}
\caption{Comparison of times to peak particle temperature $t_\mathrm{peak}$ vs. particle diameter $d_p$ for various gas oxygen concentrations for laser-ignited single iron particles at $T_\mathrm{gas} = 300$\,K. Present work (OF), \cite{mich2023} (Ref.--Sim.) and \cite{ning2021} (Ref.--Exp.) with mean and standard deviation from the measurements.}
\label{fig:validation_dp_vs_O2_vs_t}
\end{figure}

Figure\,\ref{fig:validation_dp_vs_O2_vs_t} shows a comparison of times to peak particle temperature $t_\mathrm{peak}$ (e.g. $23$\,ms in Fig.\,\ref{fig:validation_Tp_vs_t}) as a function of particle diameter and gas oxygen concentration for single laser-heated iron particles.
Increasing the particle diameter leads to extended time periods to reach the peak particle temperature as expected. Also in line with expectations, increasing the gas oxygen concentrations yields a reduction of $t_\mathrm{peak}$, e.g. by a factor of around two when increasing $X_\mathrm{O_2}$ from 21\% to 36\%. This reflects the (oxygen) diffusion-limited character of the heterogeneous reaction process of iron that is attained for the high particle temperatures of the laser-ignited particle experiments. The predictions by the present particle model implementation are fully aligned with the reference implementation from  \cite{mich2023} and both simulations agree very well with the measurements from \cite{ning2021}. As illustrated in Figs.\,\ref{fig:validation_Tp_vs_t} and \ref{fig:validation_dp_vs_O2_vs_t} the present implementation of the FOSK iron combustion sub-model is valid for a wide range of particle and environmental conditions and can therefore be used as a reliable sub-model for individual particles in the Lagrangian particle cloud within the subsequent CP-DNS of the mixing layer.

\section{Computational configuration}
\label{sec:com}

\begin{figure}[ht!]
\centering
\includegraphics[width=0.7\textwidth]{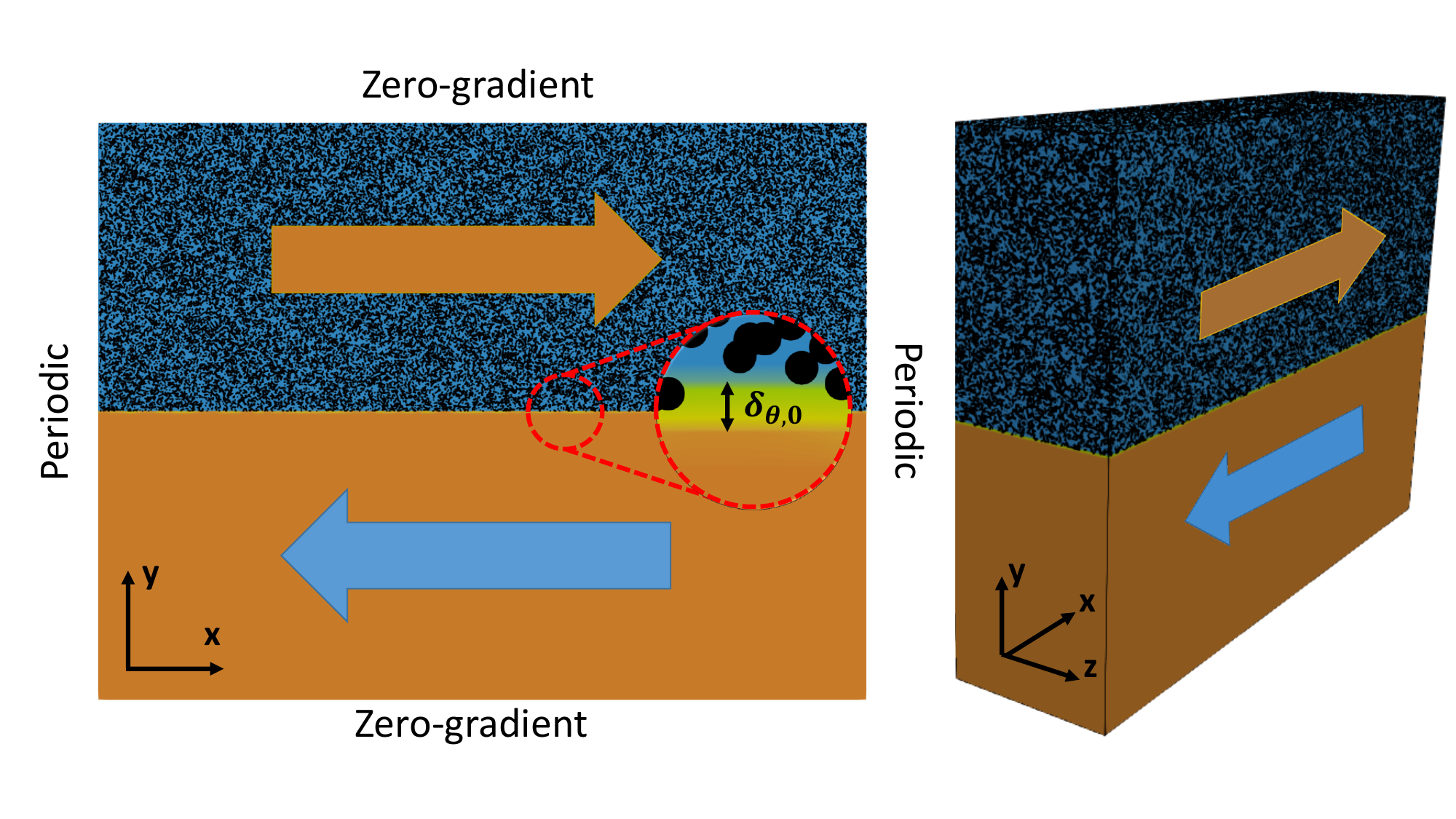}
\caption{Initial and boundary conditions of the reacting mixing layer with iron particles.}
\label{fig:conditions}
\end{figure}

A turbulent reacting mixing layer is studied, similar to the work by Rieth et al. \cite{rieth2018} on coal particle cloud combustion. However, different from the setup in \cite{rieth2018} and following the work by O'Brien et al. \cite{obrien2014} the present configuration is based on the initial definition and time evolution of the momentum thickness $\delta_{\theta}$, which allows for the evaluation of a self-similar region of the shear-induced turbulence. The setup consists of two opposed streams, initialised as air at elevated temperature ($Y_\mathrm{O_2} = 0.233$, $Y_\mathrm{N_2} = 0.767$ and $T = 550$\,K) and iron particles in the upper stream (US), and hot air at $T = 1650$\,K in the lower stream (LS). The velocity of the two streams is chosen to be equal, but directed in the opposite $x$-direction with $\Delta u_x = 30\,\mathrm{m/s}$, see Fig.\,\ref{fig:conditions}.
Following \cite{obrien2014}, a hyperbolic tangent profile is used to initialise the streamwise velocity field
\begin{equation}
u_x = \frac{\Delta u_x}{2} \mathrm{tanh} \left( \frac{y-L_y/2}{2\delta_{\theta,0}} \right)
\end{equation}
(while $u_y=u_z = 0$), where $\delta_{\theta,0}$ is the initial momentum thickness corresponding to Reynolds number
\begin{equation}
\mathrm{Re}_{\theta,0} = \frac{\Delta u_x \delta_{\theta,0}}{(\nu_\mathrm{US}+\nu_\mathrm{LS})/2} = 44.018
\end{equation}
where $\nu_\mathrm{US}$ and $\nu_\mathrm{LS}$ are the initial viscosities of the upper and lower stream, respectively. The three-dimensional computational domain has geometrical dimensions $L_x = 320\times\delta_{\theta,0}$,  $L_y = 240\times \delta_{\theta,0}$ and $L_z = 80\times \delta_{\theta,0}$. The setup consists of a total of 84,934,656 cubic Eulerian cells with a constant size of $\Delta = 100$\,\textmu m. A total of 4,900,000 spherical iron particles with an initial diameter of $d_{p,\mathrm{init}} = 10$\,\textmu m,  $T_{p,\mathrm{init}} = 550$\,K and $Y_\mathrm{Fe, init} = 1$ are randomly distributed in the upper stream. The particle diameter chosen for the simulations is close to the nominal surface mean size (12.7\,\textmu m) of the experimental PSD from \cite{fedoryk2023}. Note that these iron particles are produced via atomisation and the particle size is adjustable. In the context of the considered power plant oxidation-reduction process typical iron particle sizes are in the range 10-20\,\textmu m and therefore comparable to (the low end) of PSDs for pulverised volatile-containing solid fuels \cite{janicka2023}, while their size is expected not to change significantly after multiple oxidation-reduction cycles \cite{baigmohammadi2023}. Future simulations will use the full PSD from \cite{fedoryk2023} to explore the effects of particle size on ignition and combustion. The number of particles has been set to obtain $\phi = 1$ in the upper stream, based on the oxidation of $\mathrm{Fe}$ to $\mathrm{FeO}$ i.e. Eq. (\ref{eq:reaction}). Particle velocities are initialised with the bulk gas velocity of the upper stream. To speed up the growth of the mixing layer, isotropic velocity perturbations of $u'u' = v'v' = w'w' = 0.01 \left(\Delta u_x\right)^2$ with a length scale of $0.01 \cdot L_x$ are generated with the methods by Klein et al. \cite{klein2003} and Kempf et al. \cite{kempf2012}. These instabilities are initially superimposed on the bulk velocity field at $y/L_y = 0.5$ with a height of $4 \cdot \delta_{\theta_0}$ and exponentially decrease to zero along the $y$-axis. The boundary conditions (BC) for pressure and momentum are periodic in $x$- and $z$-direction, whereas a zero-gradient BC for momentum and ambient pressure is assumed in $y$-direction. To ensure sufficient resolution of the CP-DNS, the Kolmogorov length scale is estimated. Following Rieth et al. \cite{rieth2018} we define spatial averages for an arbitrary quantity $\Phi$ by averaging across the homogeneous $x$- and $z$-directions, i.e.
\begin{equation}
\langle \Phi \rangle (y,t) = \frac{1}{L_x L_z} \int_0^{L_x} \int_0^{L_z} \Phi(x,y,z,t) dx dz.
\end{equation}
Then, the local velocity fluctuation can be obtained as $\textbf{u}' = \textbf{u} -\langle\textbf{u}\rangle$. The average dissipation is given as $\langle \epsilon \rangle = \langle \tau_{ij} \frac{\partial u_i'}{\partial x_j} \rangle$, with the viscous stress tensor $\tau_{ij}$. Using the average dissipation rate and spatially-averaged kinematic fluid viscosity, the Kolmogorov length scale can be calculated as
\begin{equation}
\langle \eta \rangle= \left(\frac{\langle \nu \rangle^3} {\langle \epsilon \rangle}\right)^{1/4}.
\label{eq:kol}
\end{equation}
After the initial time steps of the simulation the minimum Kolmogorov length scale that can be estimated from Eq. (\ref{eq:kol}) is always larger than the computational grid size ($\eta > \Delta = 100\,\mu m$) such that all turbulent scales are resolved, as detailed further below.

\begin{figure}[h!]
\centering
\includegraphics[width=0.65\textwidth]{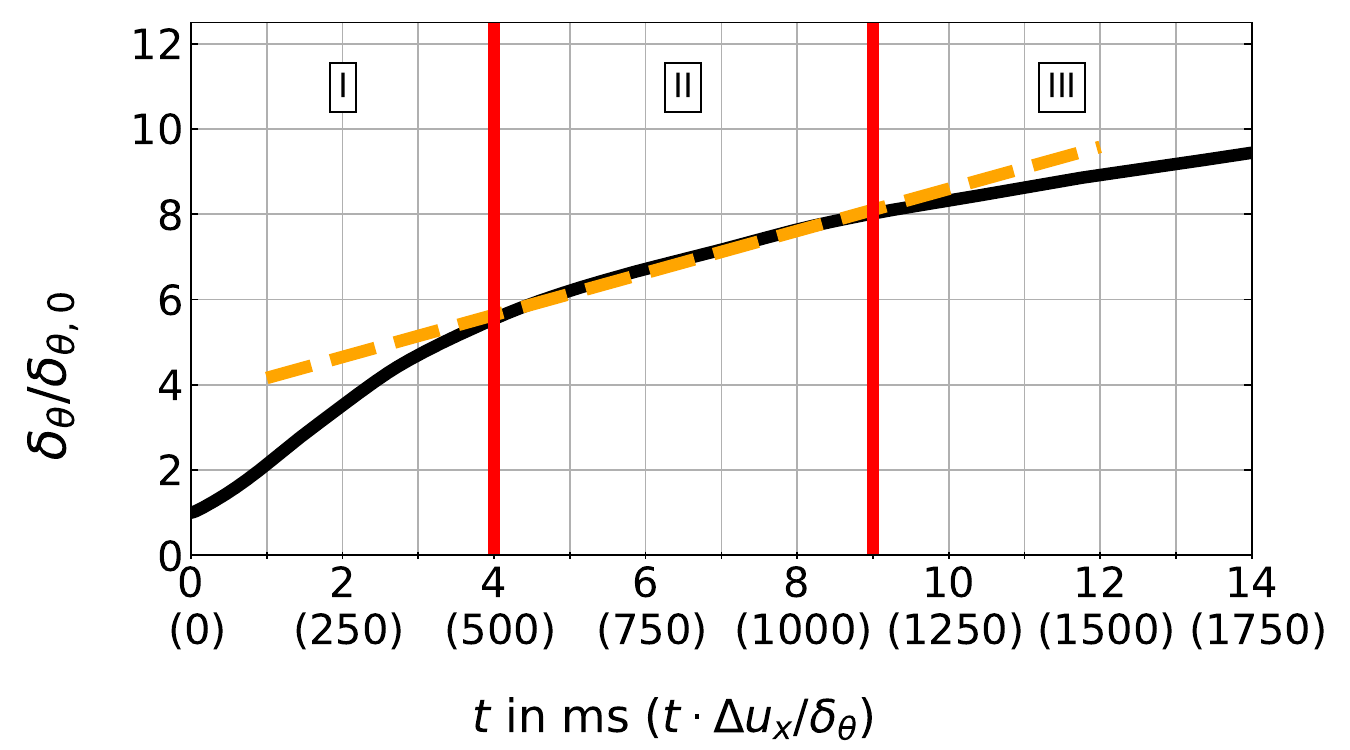}
\caption{Time evolution of momentum thickness $\delta_{\theta}$ of the mixing layer. Vertical red lines delineate the time period of self-similarity.}
\label{fig:momentum_thickness}
\end{figure}

Following the evaluation in \cite{obrien2014}, Fig.\,\ref{fig:momentum_thickness} shows the time evolution of the momentum thickness $\delta_{\theta}$ in the present mixing layer.
The momentum thickness grows with time with different ascending slopes which divide it into three different regions. The first region (I) represents the initialisation period where the initial perturbations are replaced by the development of the shear-driven turbulence. In the second region (II) a quasi-linear growth of $\delta_{\theta}$ is observed, which indicates fully-developed self-similar turbulence. The final region (III) shows a reduction of the slope which implies external processes increasingly acting on the flow field, i.e. boundary effects. Using Eq.\,(\ref{eq:kol}) a minimum Kolmogorov length scale of $213$\,\textmu m is found during the self-similar period. For the applied CP-DNS approach to be valid, the ideal grid size $\Delta_{ideal}$ should follow $d_p$\,$\ll$\,$\Delta_{ideal}$\,$<$\,$\eta$, such that all turbulent scales are resolved, while the Lagrangian particles can still be considered as point-particles. With $d_p\,=\,10\,$\textmu m, $\Delta\,=\,100\,$\textmu m and $\eta_{min}\,=\,213\,$\textmu m the present configuration achieves a reasonable trade-off during the self-similar period. The subsequent analysis will mainly focus on the self-similar region ($4$\,$<$\,$t$\,$<$\,$9$\,ms) where no boundary effects have yet affected the development of the inner shear layer. However, results from earlier and later times are also presented for a more complete documentation of the overall particle oxidation process.

The simulations are conducted using a finite volume solver based on OpenFOAM-v2012, with a typical computational cost of $\approx$165,225 CPUh on 1024 $\times$ Intel Xeon Platinum 8358 cores. Within the framework of the International Conference on Numerical Combustion ICNC (see Abdelsamie et al. \cite{abdelsamie2021}) Zirwes et al. have shown that -despite OpenFOAM's second order numerical accuracy- detailed simulations of highly-transient chemically reacting flows can be achieved that demonstrate very good HPC scaling capabilities and provide results in excellent agreement with high-order combustion codes \cite{zirwes2023}. A necessary requirement for excellent results from OpenFOAM is typically the use of a higher grid resolution than the corresponding high-order code for the same case, since the latter code will converge faster. Under this condition, depending on the mesh type, OpenFOAM has been demonstrated to provide "quasi-DNS" data for both chemically non-reacting \cite{komen2014} and reacting single phase flows \cite{zirwes2020}. Here we use an in-house reacting multiphase solver extension of OpenFOAM that has previously been employed for detailed CP-DNS analyses of volatile-containing solid fuel combustion \cite{wen2020e, wang2021, shamooni2021}. With CP-DNS, a considerable level of additional models and numerical methods are added to the classical DNS framework, which are -among others- models for the heat, mass and momentum transfer across the particle boundary layers, particle tracking techniques, particle-source in cell methods etc. To still achieve reliable CP-DNS predictions, great care has to be taken with respect to the relative size of the Lagrangian particle diameter $d_p$, Eulerian grid resolution $\Delta$ and turbulent Kolmogorov scale $\eta$, leading to the criterion $d_p$\,$\ll$\,$\Delta_{ideal}$\,$<$\,$\eta$ given above.

\section{Results and discussion} 
\label{sec:res}

After validating the iron combustion sub-model, the CP-DNS framework is used to predict the mixing layer with reacting iron particles.  
\begin{figure*}[h!]

\includegraphics[width=0.244\textwidth]{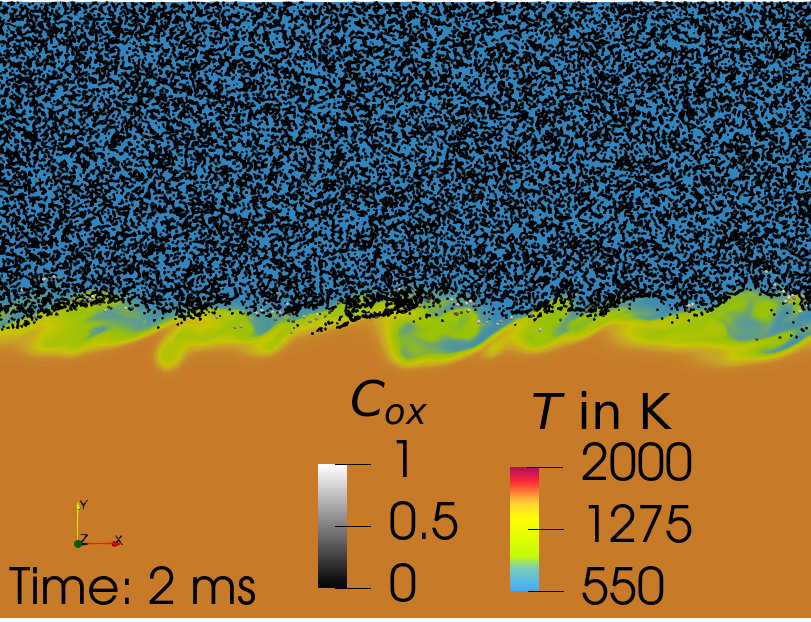}
\includegraphics[width=0.244\textwidth]{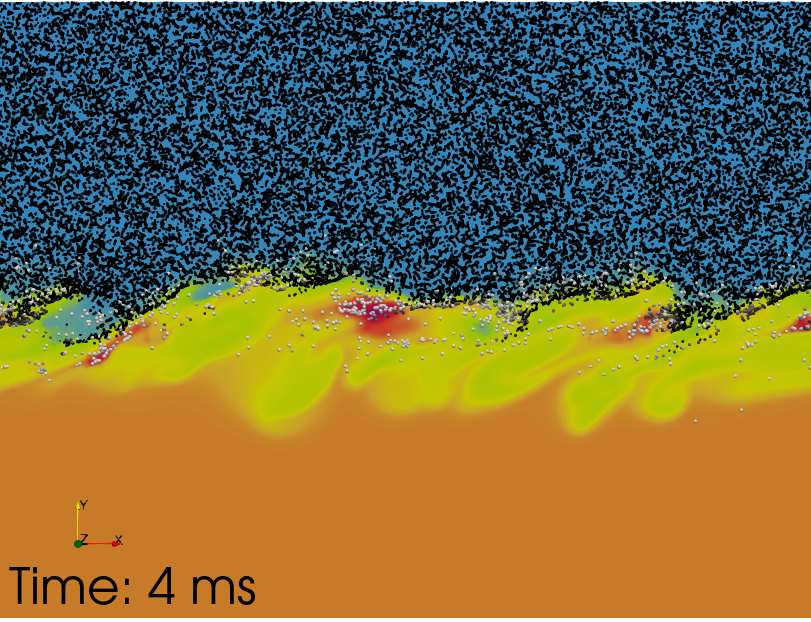}
\includegraphics[width=0.244\textwidth]{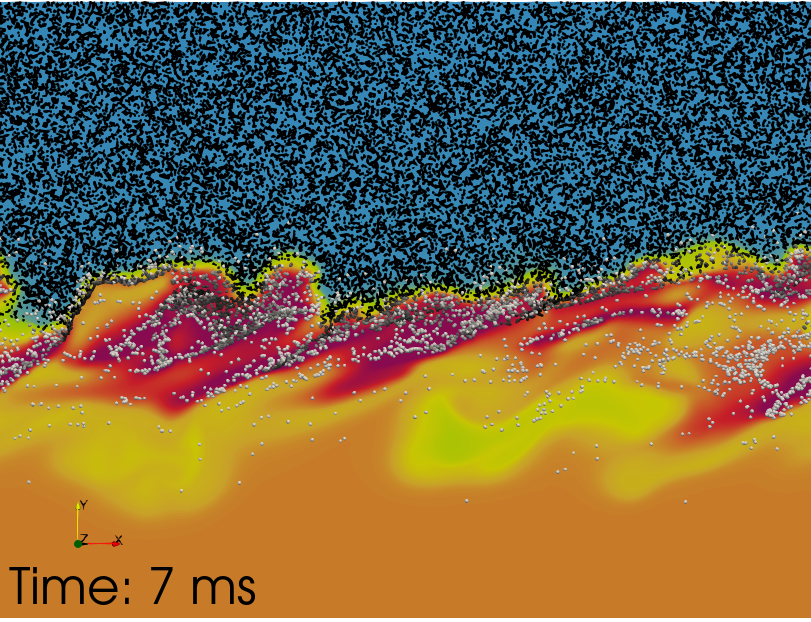}
\includegraphics[width=0.244\textwidth]{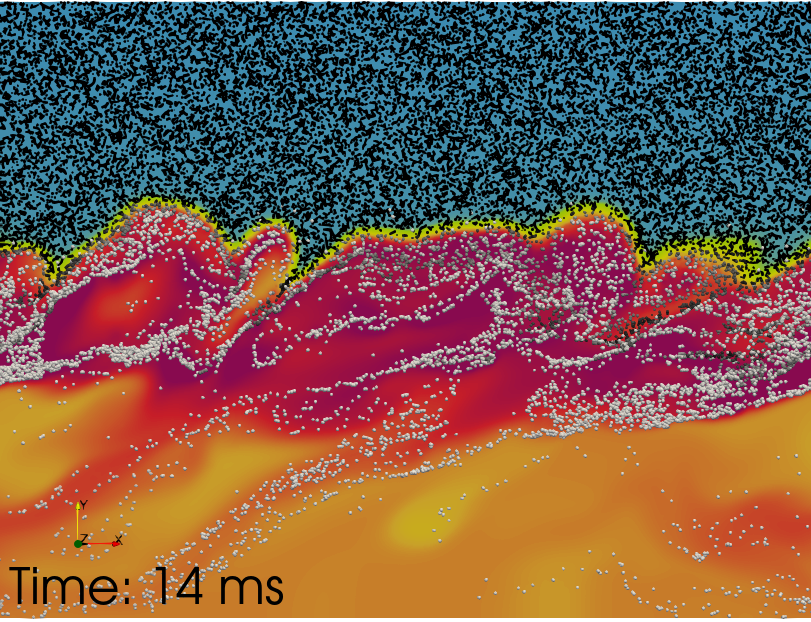}
\includegraphics[width=0.244\textwidth]{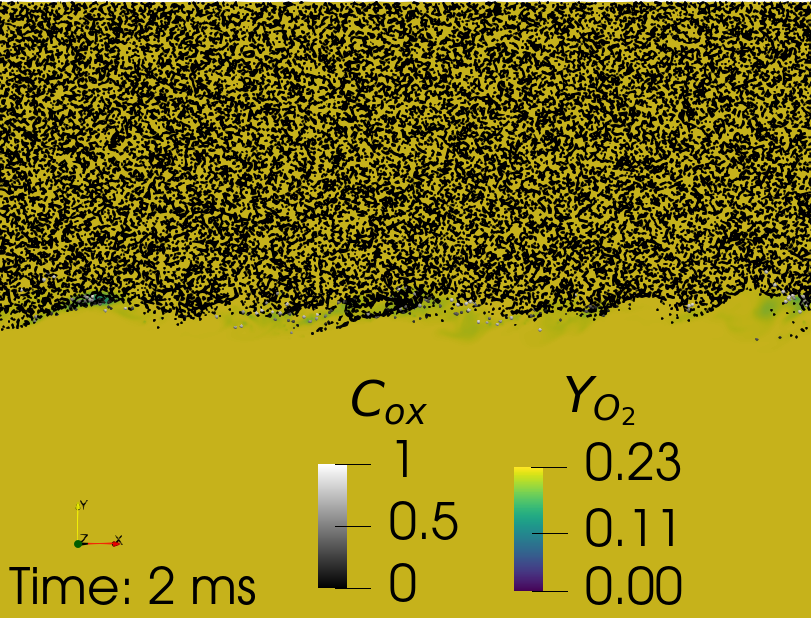}
\includegraphics[width=0.244\textwidth]{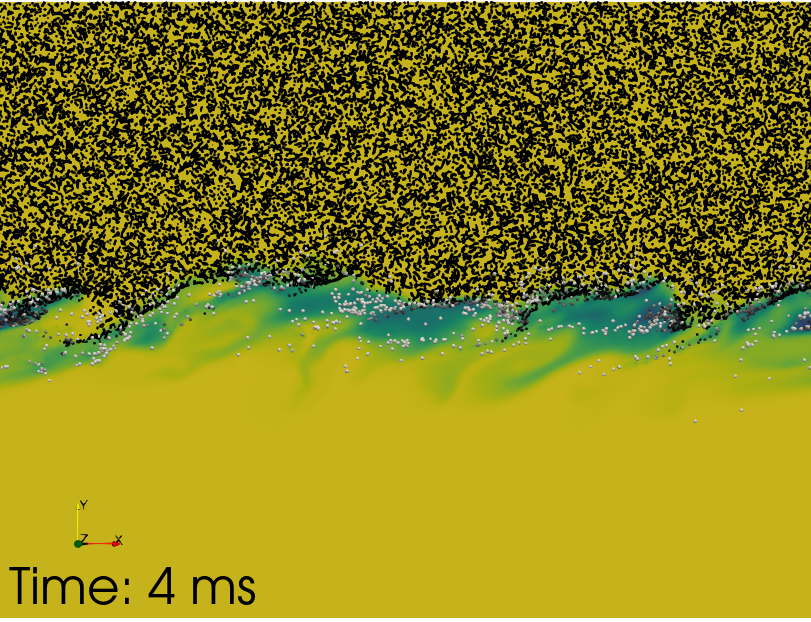}
\includegraphics[width=0.244\textwidth]{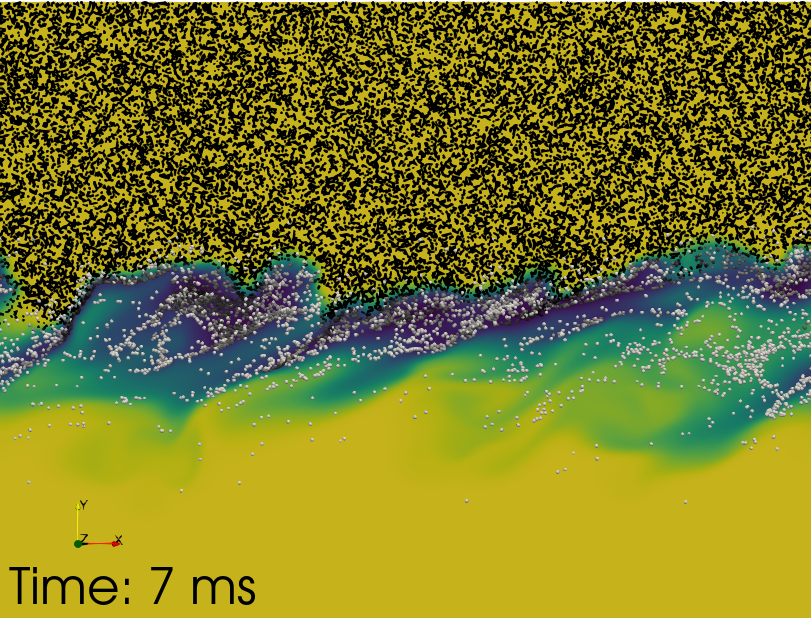}
\includegraphics[width=0.244\textwidth]{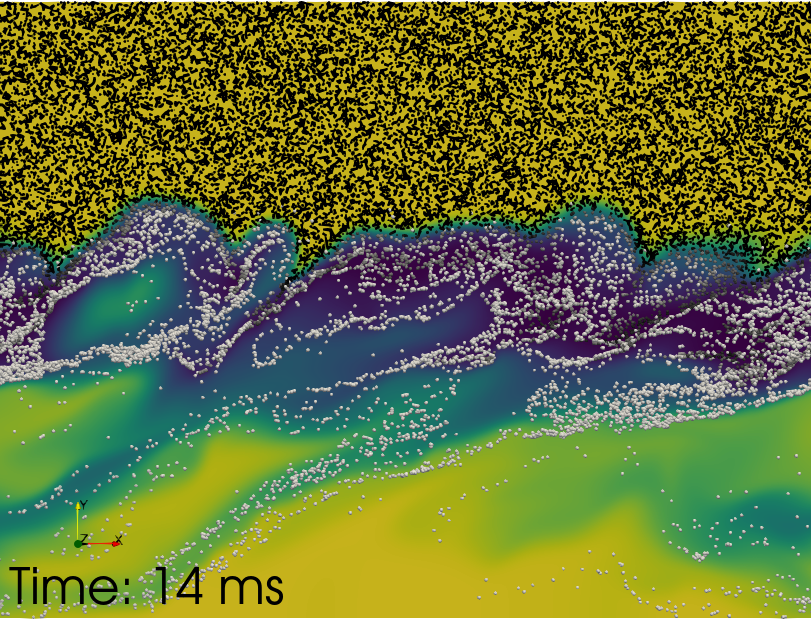}
\caption{Snapshots of gas temperature (top row) and mass fraction of $\mathrm{O_2}$ (bottom row) in the $x$-$y$ plane at $L_z/2$. The particles are coloured by their oxidation progress $C_{ox}$.}
\label{fig:mixing_layer_snapshots}
\end{figure*}
A visual impression of its temporal evolution can be obtained from Fig.\,\ref{fig:mixing_layer_snapshots} which shows snapshots of the gas temperature (top row) and oxygen mass fraction in the gas phase (bottom row) at selected times in the $x$-$y$ plane at $L_z/2$. The particles are coloured by their oxidation progress $C_{\mathrm{ox}}$ defined as the extent of iron consumption $C_{\mathrm{ox}} = \frac{m_{\mathrm{Fe}, 0} - m_{\mathrm{Fe}}}{m_{\mathrm{Fe}, 0}}$ with $m_{\mathrm{Fe}, 0}$ and $m_\mathrm{Fe}$ denoting the initial and current iron mass inside the particle at time $t$. Inspecting Fig.\,\ref{fig:mixing_layer_snapshots}, it can be seen that the initially separate streams develop and generate a mixing region between them. Due to this mixing process the particles are entrained from the upper to the lower stream, as indicated at $t = 2$\,ms. When these particles interact with the lower stream they heat up. During this initial stage of heat-up the particles are still in the kinetically-limited regime and their oxidation progress is close to zero.

At $t = 4$\,ms the turbulent mixing process has progressed and significantly more particles have been entrained into the lower stream for oxidation. Particles with longer residence times in the lower stream have reached the critical ignition temperature and ignite, as can be observed from the formation of localised hot pockets, while the local oxygen concentration reduces simultaneously. The regions of local hot pockets show the development of individual distinctive high temperature zones with highly reactive particles residing inside. However, an interesting observation is that larger pockets of substantial gas temperature increase are mostly limited to regions where particle clusters are located. 

At $t = 7$\,ms a large number of iron particles has been entrained into the lower stream, which has led to the formation of a continuous flame zone with high temperatures and full oxygen depletion. This flame zone is aligned with particle clusters in which particles typically have a similar state of oxidation. The majority of particles in the lower stream is now fully oxidised (indicated by white colour), showing that their oxidation process from Fe to FeO is fully completed. However, in the centre of the mixing layer filament-like, black-coloured particle streaks can be observed, the oxidation state of which is still zero ($C_\mathrm{ox} = 0$). Here, particle clusters have been entrained into regions with locally zero oxygen concentration, such that these particles cannot oxidise immediately. These observations differ from the well-known phenomena in volatile-containing solid fuel flames, from e.g. coal \cite{rieth2018} (Fig. 3)
and biomass \cite{rieth2018c} (Fig. 3). In high-temperature regions devoid of oxygen, volatile-containing solid fuels still continue their devolatilisation process, whereas (non-volatile) iron particles only heat up without any further material release or oxidation progress. 

For the last reported time step $t = 14$\,ms, more particles are clustered, hence, more simultaneous particle oxidation occurs and therefore, a strong increase of gas phase temperature and oxygen consumption can be observed. A continuous flame front has been established and is constantly fed by fresh particles entering from the upper stream. Nevertheless, this time step is located outside of the self-similar region, see Fig. \ref{fig:momentum_thickness}, and boundary effects may have begun to influence the results. Further snapshots at later times (omitted for brevity) show that the entrainment and oxidation progress continues until the end of the simulation.

\begin{figure}[htb!]
\centering
\includegraphics[width=0.6\textwidth]{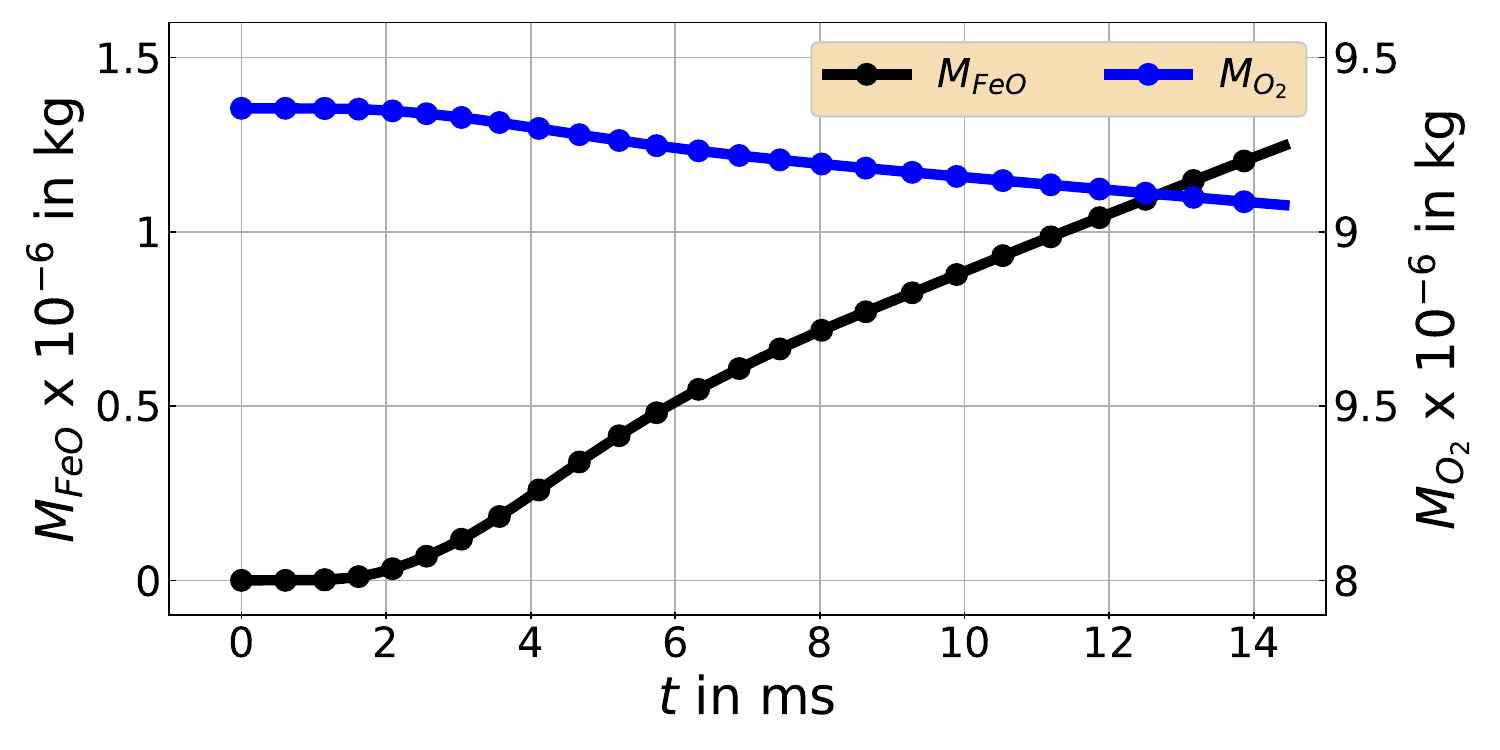}
\includegraphics[width=0.6\textwidth]{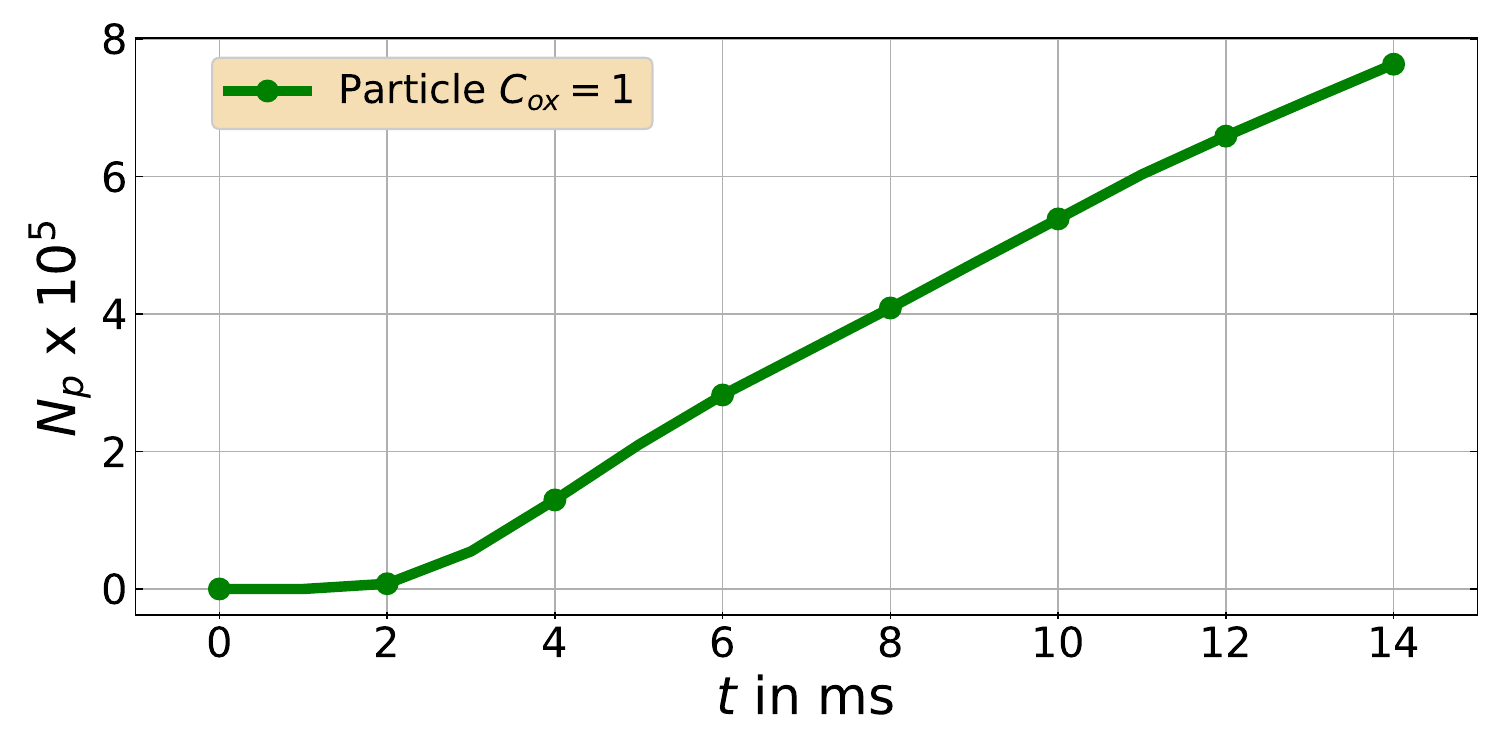}
\caption{Total mass of produced FeO and consumed O$_2$ in the entire domain (top row) and total number of fully oxidised particles $N_p$ (bottom row) vs. time.
}
\label{fig:global}
\end{figure}

Figure \ref{fig:global} (top row) shows the total mass of FeO produced and O$_2$ consumed by iron oxidation in the entire computational domain vs. time. 
A significant increase of the mass of produced iron oxide can be observed from $t = 2$\,ms onwards, after which there is a near-linear mass increase of FeO up to around 1.25 x 10$^{-6}$\,kg at $t = 14$\,ms. The mass of oxygen in the gas phase decreases simultaneously, reaching its lowest value at the end of the simulation. Figure \ref{fig:global} (bottom row) shows the total number of particles that is fully oxidised ($C_{\mathrm{ox}}$\,=\,1) as a function of time. It can be observed that $N_p$ increases simultaneously as the mass of FeO, subsequently reaches a near-linear profile and at $t = 14$\,ms around 763,000 ($\approx 15.6\,\%$) particles have been fully oxidised within the bounds of the CP-DNS simulation time.

\begin{figure}[htb!]
\centering
\includegraphics[width=0.6\textwidth]{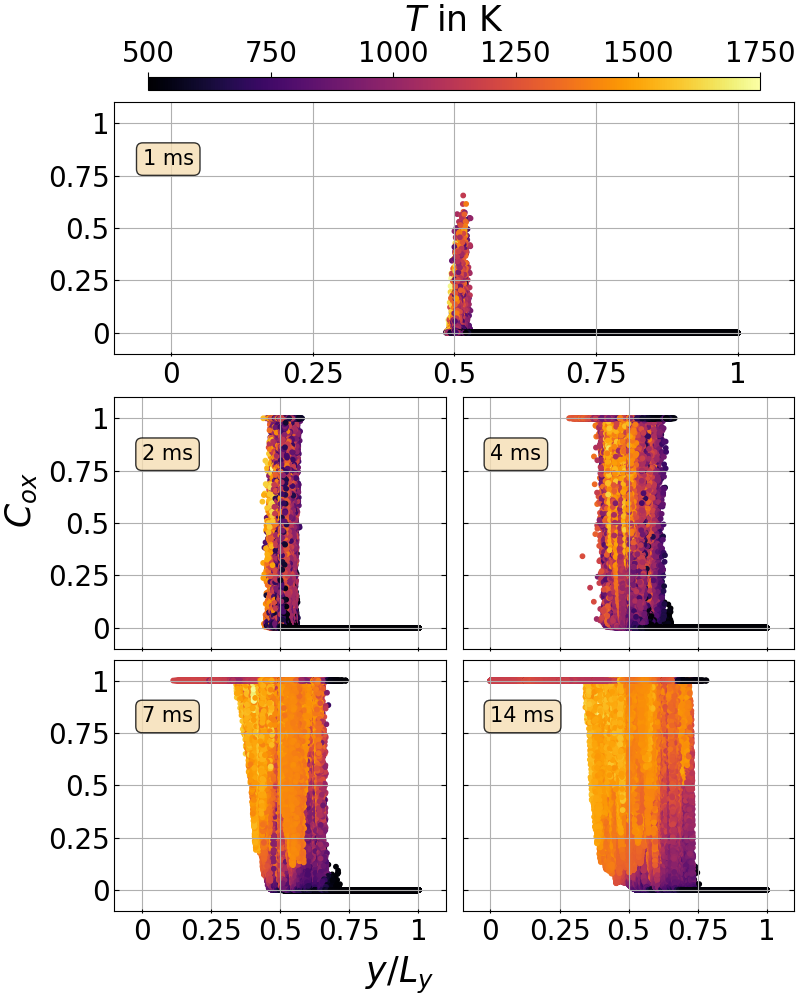}
\caption{Scatterplot of particle oxidation progress $C_\mathrm{ox}$ vs. normalised cross-stream coordinate $y$ coloured by the gas temperature surrounding the particle at different times.}
\label{fig:scatter_Cox_Y_T}
\end{figure}

Figure\,\ref{fig:scatter_Cox_Y_T} shows scatterplots of the particle oxidation progress $C_\mathrm{ox}$ vs. the normalised $y/L_y$ coordinate for all particles at different times. The particles are coloured by the gas temperature surrounding them.
At $t = 1$\,ms, a small set of particles at positions close to $y/L_y = 0.5$ has been heated up by the lower stream and begins to oxidise. As time progresses to $t = 2$\,ms, more particles are entrained into the lower stream, heat up and oxidise, such that a significant number of them has already reached the state of full oxidation $C_\mathrm{ox}\,=\,1$. This process continues up to the end of the simulation, with further broadening of the oxidation region across the $y$-axis and an increasing number of particles reaching full oxidation. At $4$\,ms it can be observed that particles in the lower half of the domain are surrounded by higher gas temperatures, whereas their surrounding gas is significantly colder in the upper half. At $y/L_y\,\approx\,0.6$ a set of particles has already reached full oxidation, despite being surrounded by relatively cold gas (black particles with $C_\mathrm{ox}\,\rightarrow\,1$). This corresponds to rare events where particles have already experienced a hot gas environment for oxidation in the lower stream and have subsequently be re-entrained upwards into to cold gas of the upper stream, see Fig.\,\ref{fig:mixing_layer_snapshots} at early times. For the time range $7$\,$\leq$\,$t$\,$\leq$\,$14$\,ms it can be observed that particles with a small oxidation progress are surrounded by cold gas, while the intermediate range of oxidation progress $0$\,$<$\,$C_\mathrm{ox}$\,$<$\,$1$ is mostly correlated with high values of surrounding gas temperature (yellow) and only a small fraction of particles is subjected to intermediate temperatures (purple). Particles that are fully oxidised ($C_\mathrm{ox}\,=\,1$) can be observed across the entire range of $y/L_y$ of the lower stream and up to about half of the upper stream at $t = 14$\,ms. These particles are either surrounded by the background gas temperature of the lower stream (orange), are subjected to the hot gases of the main oxidation region (yellow) for intermediate values of $y/L_y$, or surrounded by the low gas temperatures of the upper stream (purple-black), where -again- the latter correspond to rare re-entrainment events of oxidised particles into the upper stream. The fraction of fully reacted particles re-entrained into the cold gas is less than 5\% for the inspected time range. These particles transfer heat to the cold gas, but due to their small fraction, no significant effect on the gas phase can be observed, see gas temperature in Fig. 5. We note that for volatile-containing solid fuels that undergo char conversion (small) particles would be fully consumed at (very) late times of the simulation, whereas this cannot happen for iron particle combustion where, apart from few particles having left the computational domain through the y-boundaries, no significant particle loss has occurred at the end of the simulation.

\begin{figure}[htb!]
\centering
\includegraphics[width=0.6\textwidth]{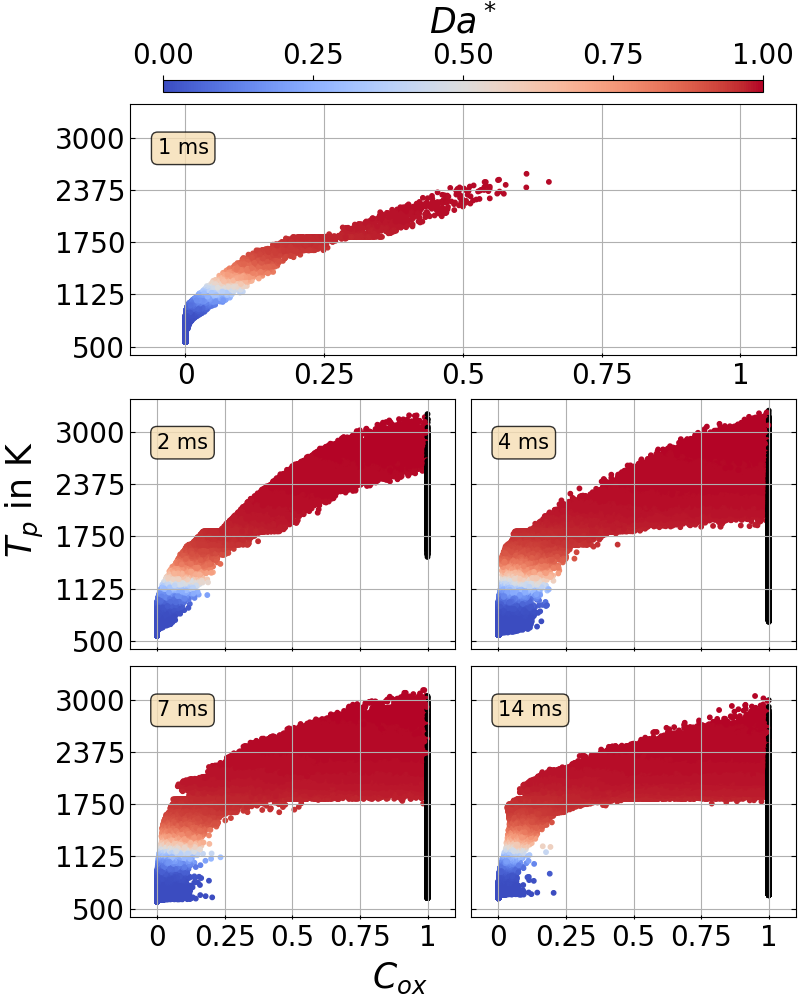}
\caption{Scatterplot of particle temperature $T_p$ vs. oxidation progress $C_\mathrm{ox}$ coloured by the normalised Damköhler number $\mathrm{Da}^*$ at different times. Black colour indicates a fully oxidised state.}
\label{fig:scatter_Tp_Cox_Da}
\end{figure}

Figure\,\ref{fig:scatter_Tp_Cox_Da} presents scatterplots of particle temperature vs. oxidation progress $C_\mathrm{ox}$ over the course of the simulation, where particles are coloured according to their normalised Damköhler number $\mathrm{Da}^*$ (Eq. \ref{eq:damkohler}). Fully oxidised particles are coloured in black since their mass conversion rate of Fe to FeO is zero and the evaluation of $\mathrm{Da}^*$ becomes obsolete. In line with Fig.\,\ref{fig:scatter_Cox_Y_T} at $t = 1$\,ms a first set of particles has started to partially oxidise. 
Corresponding to the second plateau in Fig.\,\ref{fig:validation_Tp_vs_t}, a constant maximum particle temperature across the range $0.2$\,$\leq$\,$C_{ox}$\,$\leq$\,$0.25$ delineates the melting process of Fe at $T_p = 1810$\,K. At $t\,=\,2$\,ms some particles have already reached $C_\mathrm{ox}\,=\,1$, where the highest particle temperatures are attained. Considering the reactants solid iron at the initial temperature of the upper stream ($T=550$\,K) and gaseous air at the temperature of the lower stream ($T=1650$\,K), while assuming the products of iron oxidation to be liquid FeO and gaseous nitrogen only, an adiabatic flame temperature of more than 3100\,K is obtained, in line with the peak temperatures observed in Fig.\,\ref{fig:scatter_Tp_Cox_Da}. At $C_\mathrm{ox}\,=\,1$ (black particles) particle oxidation has completely finished such that particles cool down again to a level determined by their local gas environment. For the time range $4$\,$\leq$\,$t$\,$\leq$\,$7$\,ms the particle scatter becomes substantially thicker due to many particles attaining a broad range of oxidation states and temperatures. These particles are mainly located in the main flame region in the centre of the mixing layer, see Fig.\,\ref{fig:scatter_Cox_Y_T}. At $t\,=\,14$\,ms, the temperature range for particles with elevated oxidation progress ($C_\mathrm{ox}\,>\,0.3$) is $1750$\,$\leq$\,$T_p$\,$\leq$\,$2900$\,K, which corresponds to the set of particles in the main reaction zone in Fig.\,\ref{fig:mixing_layer_snapshots}. Focusing on the colour of the particles, i.e. the normalised Damköhler number, it can be observed that for $T_p$\,$<$\,$1125$\,K and $C_\mathrm{ox}$\,$<$\,$0.2$, all particles are in the kinetically-limited regime ($\mathrm{Da}^* \rightarrow 0$) as expected. For $1125$\,$<$\,$T_p$\,$<$\,$1500$\,K, the transition from kinetic to diffusion limitation ($\mathrm{Da}^* \rightarrow 1$) occurs. This is because particle temperatures are so high such that the conversion rate is now solely limited by the availability of oxygen. 

In the previous figures the (strong) impact of turbulence on iron particle ignition and combustion in the mixing layer has been demonstrated. 
\begin{figure}[h!]
\centering
\includegraphics[width=0.65\textwidth]{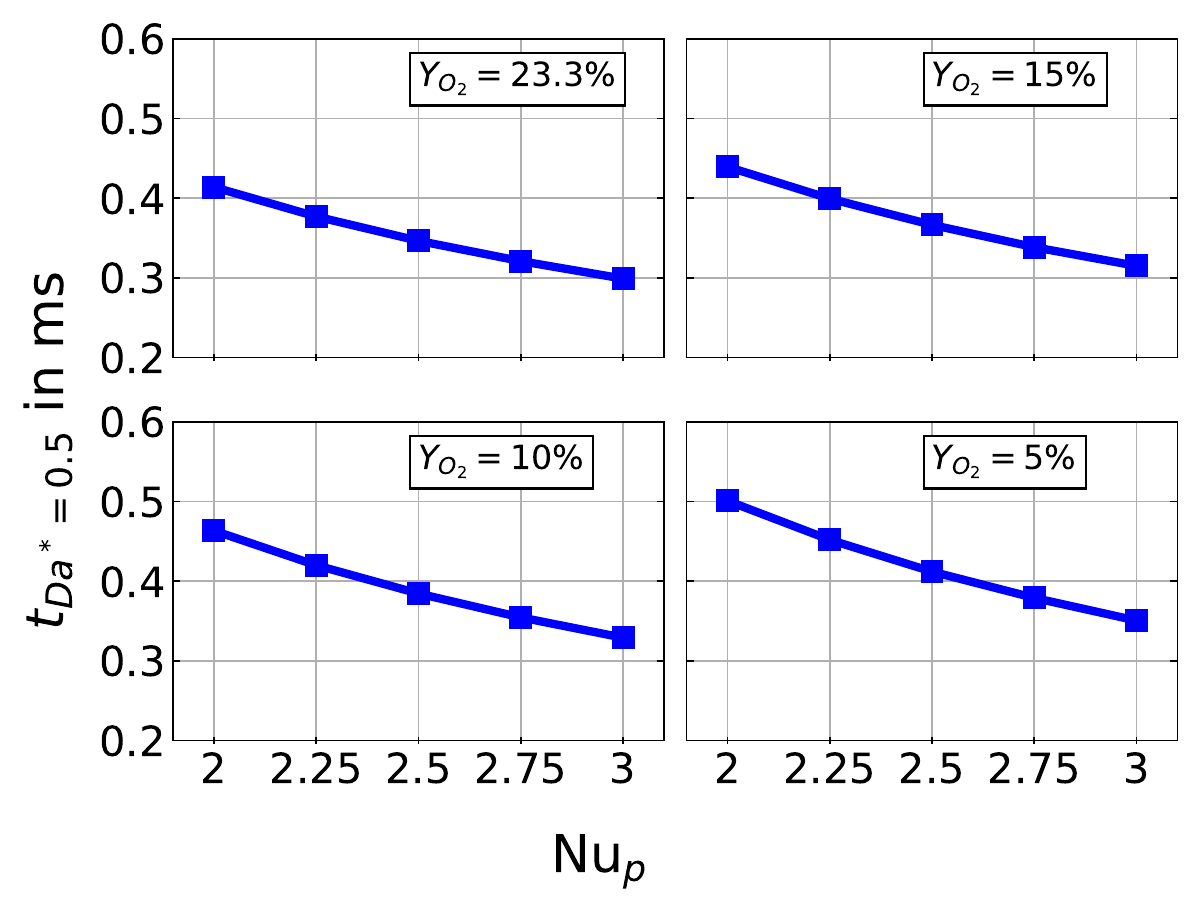}
\caption{Time to normalised Damköhler at transition point $t_\mathrm{Da^*=0.5}$ vs. particle Nusselt number Nu$_p$ for various gas oxygen environments for a single iron particle at $T_p = 550$\,K in $T_\mathrm{gas} = 1650$\,K.}
\label{fig:t_vs_Nu_O2}
\end{figure}
For a quantitative evaluation of turbulence effects on the transition from kinetic to diffusion limitation of particle conversion Fig.\,\ref{fig:t_vs_Nu_O2} shows the time to reach this transition ($\mathrm{Da}^* = 0.5$) for different particle Nusselt numbers in varying oxygen environments. The data shown in Fig.\,\ref{fig:t_vs_Nu_O2} are generated by conducting separate simulations of single iron particle conversion in various oxygen environments using the same models and implementation from Sec.\,\ref{sec:mod}. The conditions are chosen to mimic entrainment events from the mixing layer with initially cold particles at $T_p = 550$\,K entering the lower stream at $T_\mathrm{gas} = 1650$\,K. The range of particle Nusselt numbers has been extracted from the mixing layer during the self-similar period ($t\,=\,4\,-\,9$\,ms). Here, Nu$_p = 2$ indicates a particle that perfectly follows the flow field (zero slip velocity), see Eq.\,\eqref{eq:Nu}, whereas for Nu$_p = 3$, a slip velocity between the particle and gas is present that is enhanced by turbulent mixing. In Fig.\,\ref{fig:t_vs_Nu_O2} a decrease of oxygen in the environment leads to an increase of transition time because of the direct dependence of the FeO formation rate on oxygen concentration, see Eq.\,\eqref{eq:mFeO}. Increasing the particle Nusselt number, equivalent to increasing the degree of turbulence, leads to a $\approx 25-30$\,\% faster process of oxidation.

In summary, our results show some similarities, but also distinct differences between non-volatile iron and volatile-containing solid fuel flames in the present shear-driven turbulence, namely remaining oxide particles at late times and a stronger dependence of particle conversion on the local availability of oxygen for iron combustion.

\section{Conclusions}
\label{sec:con}

CP-DNS of iron particle cloud ignition and combustion in a turbulent mixing layer is conducted using existing sub-models for iron particle combustion \cite{mich2023}. The iron combustion sub-model is successfully validated against single-particle reference data and capable of recovering measured particle temperatures, melting and solidification phenomena. Subsequently, the model is used for CP-DNS of iron particle cloud ignition and combustion in shear-driven turbulence. Simulation results show that particles are entrained into the hot lower stream, where ignition occurs and the heat release from particle oxidation increases the gas temperature far above the background temperature. While the local oxidation of individual particles has a limited effect on gas temperature, the major heat transfer to the gas phase originates from particle clusters. The global analysis of FeO mass production and O$_2$ consumption from oxidation across the entire domain shows that FeO is continuously produced, while molecular oxygen is consumed, leading to localised regions fully depleted of oxygen at late times. The analysis of the normalised Damköhler number, describing the transition from kinetically-limited to diffusion-limited iron combustion, shows that kinetic limitation applies for particle temperatures and particle oxidation progress of less than 1125K and 20\,\%, respectively. For higher particle temperatures the oxidation rate is diffusion-limited and peak particle temperatures are observed near the fully-oxidised particle state. The major differences between the present non-volatile iron flames and volatile-containing solid fuel flames are non-vanishing particles at late simulation times and a stronger limiting effect of the local oxygen concentration on the overall conversion process in iron dust flames. Future work will remove the present assumption of mono-sized iron particles and explore the effects of polydispersity on iron particle cloud ignition and combustion.

\section*{Acknowledgement}

This work is conducted within the \textit{Clean Circles} research initiative financially supported by KIT Strategiefonds and the Hessian Ministry of Higher Eduction, Research, Science and the Arts. O.T. Stein gratefully acknowledges support by the Helmholtz Association of German Research Centres (HGF), within the research field \textit{Energy}, program \textit{Materials and Technologies for the Energy Transition (MTET)}. The authors acknowledge support by the state of Baden-Württemberg through bwHPC and are grateful for HPC time on HLRS Hawk.

\bibliographystyle{ieeetr}
\bibliography{bibliography}

\end{document}